\documentstyle[epsfig]{mn}
\def\msun{\mbox{M$_\odot$}}
\def\I{{\'\i}}
\def\HII{H\,{\sc ii} }

\title[PN-Carbon Yields and the Chemical Evolution of the Galaxy]
{PN-Carbon Yields and the Chemical Evolution of the Galactic Disk}
\author[L. Carigi]
{Leticia Carigi \\  
Instituto de Astronom\'{\i}a, Universidad Nacional Aut\'onoma de M\'exico, 
A.P. 70-264, 04510 M\'exico, D.F.  }

\date{\today}
 
\begin{document}
\maketitle

\begin{abstract}

Two sets of observational carbon stellar yields for low-and-intermediate mass stars 
are computed based on
 planetary nebula abundances  derived from C II $\lambda4267$ and 
C III $\lambda\lambda1906+1909$ lines, respectively.
These observational yields are assumed in 
chemical evolution models for the solar vicinity and the Galactic disk.
C/O values observed in stars of the solar vicinity
and Galactic \HII regions are compared with those predicted by chemical evolution models 
for the Galaxy.
I conclude that the C yields derived from permitted lines are in better
agreement with the observational constraints
than those derived from forbidden lines. 

\end{abstract}
 

\section{Introduction}
\label{sec:intro}

The $N({\rm C}^{++})/N({\rm H}^+)$ values derived from 
the optical recombination line (permitted line, PL) C II $\lambda4267$ are higher, 
by as much as a factor of 10, 
than those determined from the collisionally excited lines (forbidden lines, FL) 
C III $\lambda \lambda1906+1909$
(eg. Rola \& Stasinska 1994, Peimbert, Luridiana \& Torres-Peimbert 1995a,
Peimbert, Torres-Peimbert \&  Luridiana 1995b, Liu et al. 2001, Luo, Liu \& Barlow 2001).
Several explanations for this discrepancy have been presented in the literature,
(see the reviews by Liu 2002, Peimbert 2002, Torres-Peimbert \& Peimbert 2002)
but the problem remains open.
Since PNe are important for the C enrichment of the interstellar medium, 
a successful chemical evolution model for the solar vicinity and the Galactic
disk (Carigi 2000) is used 
to discriminate between the PN-C abundances derived from permitted lines
and the PN-C abundances obtained from forbidden lines. 
Hereafter, all abundances are given  by number.

This work is based on a preliminary study presented by Carigi (2002).

\section{Observational Constraints}
In this work, the data used as observational constraints are the following:
i) C and O abundances from Galactic \HII regions and Galactic B-stars to constrain the present-day abundance gradient,
ii)  C and O abundances from different objects in the solar vicinity to constrain the  C/O history.
The observational constraints are presented in Figure 1.

The new C/H and O/H gaseous values for the three \HII regions M17, M8 and Orion,
(at $r=$ 5.9, 6.5 and 8.4 kpc,
adopting the  Galactocentric distance for the Sun of 8 kpc)
are taken from Esteban et al. (2002).
These values have been increased 0.10 dex and 0.08 dex, 
respectively, due to the fraction
of these elements embedded in dust grains (Esteban et al. 1998).
The C/H and  O/H abundances are derived from
the permitted lines C II $\lambda 4267$ and O II $\lambda 4649$, respectively.
Esteban et al. (2002) re-calculated the C/H values given by Esteban et al. (1998, 1999a, 1999b),
(mainly an error in C/O of M17 was corrected)
and these values are presented in their Table 10.
Based on these \HII regions the C/H, O/H and C/O  gradients are
$-0.086$, $-0.049$, and $-0.037$   dex kpc$^{-1}$, respectively.

The C/H and O/H values for B-stars are taken from Rolleston et al. (2000) and Smartt et al. (2001).
In Figure 1 (a)--(c) I show only the values for those stars that have both C and O determinations.
Galactic distances have been adjusted to $r_\odot=8$ kpc.
The C/H, O/H and C/O gradients from B-stars are 
$-0.07\pm$0.02, $-0.067\pm$0.008, and $-0.05 \pm$ 0.02 dex kpc$^{-1}$,
respectively.

The C/H and C/O gradients  from B-stars are in agreement with those from \HII regions,
but the C/H values from  B-stars are lower by 0.5--1 dex than those determined from \HII regions.
The O/H values from \HII regions and B-stars are similar, but
the O/H gradient for B-stars is steeper than that obtained from \HII regions by a factor of 
1.4.
There is no difference between the O/H gradient computed for 4--10 kpc and that determined by 
Smartt et al. (2002) for 2--17 kpc.
The computed C/O gradient for 4--10 kpc is closer to that obtained from \HII regions than
to that determined from B-stars located between  2--17 kpc.

Since in the literature there are C values based on recombination lines only for 
M17, M8 and Orion,
I pay attention to the chemical evolution of the Galactic disk only for 4 kpc $<r<$ 10 kpc.

The observed rise of C/O with time or metallicity in the solar vicinity is indicated by
dwarf stars located closer than 1 kpc around the Sun (Gustafsson et al. 1999) and 
by the average solar value from 
 Allende-Prieto et al. (2001, 2002) and Holwerger (2001).
Average value for two B-stars, one in  NGC 3292 and the other in Cepheus OBIII 
(two galactic clusters close to the Sun),
is also shown, despite the fact that
 they have C/O abundances lower than Orion by 0.91 and 0.98 dex, respectively.

\begin{figure}
\epsfxsize=9cm
\epsfbox{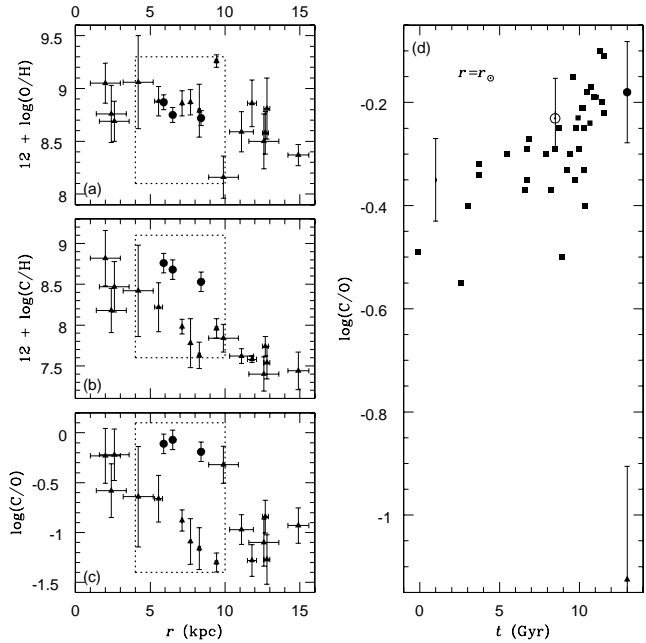} 
{\caption[]{Observational constraints for the models.
(a)--(c) Present-day distribution of abundance ratios.
Only the data of the enclosed area will be assumed as observational constraints.
{\it Filled circles:} \HII regions,  gas and dust values from
Esteban et al. (2002, 1998).
The observed values correspond to M17, M8, and Orion at adopted Galactocentric distances
of 5.9, 6.5, and 8.4 kpc, respectively.
{\it Filled triangles:} B-stars from Rolleston et  al. (2000) and  Smartt et al. (2001),
only from stars with both C and O determinations.
(d) The C/O evolution of the solar vicinity.
{\it Filled circle:} Orion from Esteban et al. (2002, 1998).
{\it Filled triangle:} average value for the two B-stars at $r=r_\odot \pm 0.5$ kpc
from data by Rolleston et  al. (2000).
{\it Filled squares:} dwarf stars at $r=r_\odot \pm 1$ kpc from Gustafsson et al. (1999).
The ages of the dwarf stars were scaled to the age of the models.
Error bar at the left represents the typical error.
{$\odot$:} average solar value from
Allende-Prieto et al. (2001, 2002) and Holweger (2001).
}
\label{f1}}
\end{figure}

\section{PN Observational Yields} 
\label{sec:constraints}

\begin{figure}
\epsfxsize=9cm
\epsfbox{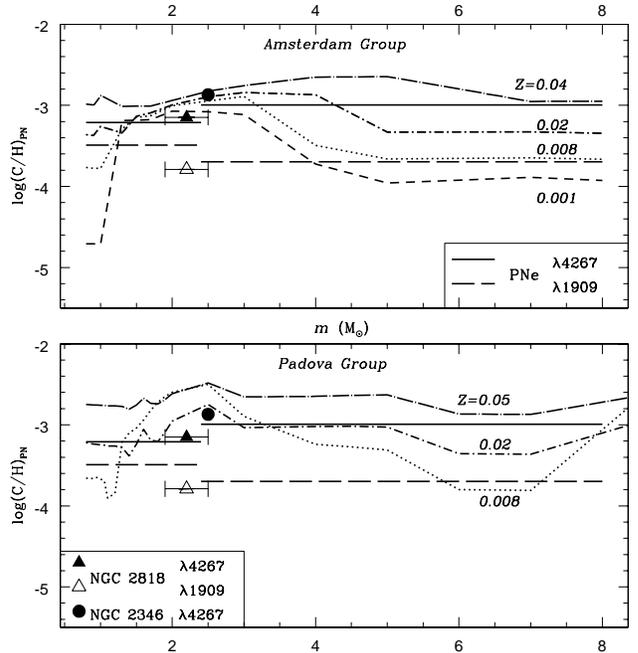} 
{\caption[]{
Log(C/H) in Planetary Nebulae vs initial mass of the PN progenitors.
Predictions from
Amsterdam yields (van den Hoek \& Groenewegen 1997) and from Padova yields
(Marigo et al. 1996,  1998, Portinari et al. 1998) for different
initial metallicities.
The horizontal lines represent observational
average values computed using the permitted line C II $\lambda 4267$
and forbidden lines CIII $\lambda\lambda 1906+1909$ from
type I PNe ($ m > 2.4$ \msun) (Peimbert et al. 1995a)
and  type II and III PNe ($ m < 2.4 $ \msun) (Peimbert et al. 1995b).
The observational data for NGC 2818 and NGC 2346 come from Dufour (1984) and 
Peimbert \& Serrano (1980), respectively.
}
\label{f2}}
\end{figure}

Based on the classification by Peimbert (1978),
in this work I assume that 
type I PN progenitors are stars with initial mass between 2.4 and 8 M$_\odot$, 
and type II and III PN  progenitors are stars with initial mass between 0.8 and 2.4 M$_\odot$.
The C yields for PN progenitors 
are calculated based on the average (C/H)$_{\rm PN}$ 
from permitted lines (C$_{\rm PN}^{\rm PL}$ yields)
 and forbidden lines (C$_{\rm PN}^{\rm FL}$ yields),
neglecting the ejected mass by winds and assuming that  the  average (C/H)$_{\rm PN}$ values 
are independent of the initial metallicity and mass of the progenitors. 

Based on the previous assumptions, 
the total ejected mass ($m_e$) is the same as the mass ejected by the PN event and
 the stellar yield of the element $j$,
 $p_j$,  can be written as

$$ p_j \sim m_e/m (X_j^{\rm PN} - X_j^i), $$

\noindent
where 
$m$ is the mass of the PN progenitor in the main sequence,
$X_j^{\rm PN}$ is the abundance by mass of the element $j$ determined in 
the planetary nebula, 
and $X_j^i$ is the initial stellar  abundance  or the abundance
of the molecular nebula
where the PN progenitor was formed.  
Therefore, the C yield for PN progenitors as a function of the observed average (C/H) ratio is

$$p_C^{\rm PN} =12 \left( <{\rm(C/H)}^{\rm PN}> (p_H + {\rm H}^i m_e/m ) - {\rm C}^i m_e/m \right). $$

The $m_e$ values and the corresponding H yields for PN progenitors
are taken from  van den Hoek \& Groenewegen (1997), while ${\rm H}^i$ and ${\rm C}^i$
are computed by the chemical evolution code.
Since $<{\rm (C/H)}_{\rm PN}>$ is independent of the PN progenitor metallicity,
$p_C^{\rm PN}$ depends only  on the initial stellar $Z$ through $p_H$, $m_e$,
${\rm H}^i$ and ${\rm C}^i$; this dependence is very weak. 
Moreover, $p_C^{\rm PN}$ is less dependent than the theoretical yields on the initial stellar mass,
because $<{\rm (C/H)}_{\rm PN}>$ is assumed to have a unique value for PNI progenitors and another
value for PNII/III ones.
I assume that the O yields for PN progenitors are null.

Average (C/H)$_{\rm PN}^{\rm PL}$ and (C/H)$_{\rm PN}^{\rm FL}$ values are computed from
15 type I PNe (PNI), and from 21 type II and III PNe (PNII/III).
The average (C/H)$_{\rm PNI}^{\rm PL}$ is calculated from the $N({\rm C}^{++})/N({\rm O}^{++}$)
and $N({\rm O})/N({\rm H})$ values
given by Peimbert et al.  (1995a).
The average (C/H)$_{\rm PNI}^{\rm FL}$ value
is obtained from the average (C/H)$_{\rm PNI}^{\rm PL}$ and the average
of $N({\rm C}^{++})/N({\rm H}^+)_{\rm PL}/N({\rm C}^{++})/N({\rm H}^+)_{\rm FL}$
ratios taken from Peimbert et al. (1995b).
The $<$(C/H)$_{\rm PNII/III}^{\rm PL}>$  and  $<$(C/H)$_{\rm PNII/III}^{\rm FL}>$ values are computed from
the $N({\rm C}^{++})/N({\rm H}^+)$ given by Peimbert et al. (1995b) and corrected for the contribution of
$N({\rm C}^+)/N({\rm H}^+)$ by adding 0.1 dex.
These average values are shown in Figure 2.
In addition, I present for comparison in Figure 2  
C/H ratios  computed from theoretical yields (TY),  
assuming that the wind contribution to the yield is null,

$${\rm (C/H)}_{PN}^{TY} = (p_C/12 + {\rm C}^i m_e) / (p_H + {\rm H}^i m_e). $$

The assumed theoretical yields are
by van den Hoek \& Groenewegen (1997)
(Amsterdam yields)  and by
Marigo et al. (1996, 1998) and Portinari et al. (1998)
(Padova yields).

As can be noted from Figure 2,
the ${\rm (C/H)}_{PN}$ values derived from the theoretical yields depend on the initial stellar mass
and metallicity. 

From the observed ${\rm (C/H)}_{PN}$ values I have derived observational yields for the 0.8 to 8 \msun range.
Those yields derived from the permitted lines are in good agreement with the theoretical yields for
high $Z$ values, while the yields from forbidden lines are smaller that the theoretical ones
for high $Z$ values and agree with the theoretical ones for $Z < 0.008$.
The observational yields have been used to compute chemical evolution models.

Note that due to the small number of observational points the adopted yields are independent of $t$
and consequently independent of $Z$.
This assumptions will be tested.
Models based on the observational yields will be confronted with the observations, and as I will show
further on, it is necessary to assume models where the yields increase with metallicity to
obtain a very good agreement with the observations.

It has been possible to determine the initial mass of the stellar progenitor for
two planetary nebulae:
NGC 2818 and NGC 2346, which belong to a globular cluster and a binary system, respectively
(Dufour 1984, Peimbert \& Serrano 1980).
Their C/H values and initial masses are shown in  Figure 2.

According to Liu et al. (2000, 2001), Luo et al. (2001) and Pequignot et al. (2002),
the large discrepancy between (C/H)$^{\rm PL}$  and (C/H)$^{\rm FL}$
may be caused by cold and extra-metal-rich condensations
of very low mass and high density, embedded in the hot and diluted material of PNe.
The C II $\lambda$ 4267 is emitted in these cold regions and the PNe abundances 
determined from this line are not representative of the ejected material. If this
idea is correct, the  C$_{\rm PN}^{\rm FL}$ yields are more realistic than the C$_{\rm PN}^{\rm PL}$ yields.
 
According to Torres-Peimbert \& Peimbert (2002) in addition to the idea of chemical inhomogeneities,
there are six other possible causes for the observed differences between the  abundances derived from
FL and PL: shadowed regions, density variations, deposition of mechanical energy, deposition of
magnetic energy, dust heating, and decrease of the ionizing flux with time.
If a combination of these six mechanisms is responsible of the abundance differences,
then the  C$_{\rm PN}^{\rm PL}$ yields are more realistic  than the C$_{\rm PN}^{\rm FL}$ ones.

\section{Models }
\label{sec:models}

\begin{table}
\caption{Present-day radial Gradients}
\begin{center}
\begin{tabular}{lccc}
\multicolumn{1} {c} {Assumed yields} & \multicolumn{3} {c}{Gradients $^a$ (dex kpc$^{-1}$)} \\
MS+LIMS  & C/H & O/H & C/O \\
\hline
\hline
{Geneva +} & & & \\
Padova             & -0.084 & -0.048 & -0.036 \\
Amsterdam          & -0.098 & -0.049 & -0.049 \\
PNe $\lambda4267$  & -0.091 & -0.048 & -0.042 \\
PNe $\lambda1909$  & -0.096 & -0.049 & -0.047 \\
\hline
{Padova +} & & & \\
Padova             & -0.056 & -0.053 & -0.003 \\
PNe $\lambda4267$  & -0.063 & -0.054 & -0.009\\
PNe $\lambda1909$  & -0.067 & -0.055 & -0.012\\
\hline
\multicolumn{1}{c} {Observations}  & & & \\
\HII regions$^b$             & -0.086 & -0.049 & -0.037 \\
B-stars$^c$                 &-0.07$\pm$0.02 & -0.07$\pm$0.01 & -0.05 $\pm$0.02 \\
B-stars$^d$                 &-0.105 & -0.067 & -0.038  \\
\hline
\hline
\end{tabular}
\end{center}
\begin{flushleft}
$^a$ Average value of two predicted gradients, that between 4--10 kpc and that between 6--8 kpc.

$^b$ Computed from Esteban et al. (1998, 1999, 2002) 

$^c$ Smartt et al. (2001) between 2 and 17 kpc

$^d$ Computed from Rolleston et al. (2000) between 4 and 10 kpc
\end{flushleft}
\end{table}

\begin{figure}
\epsfxsize=9.0cm
\epsfbox{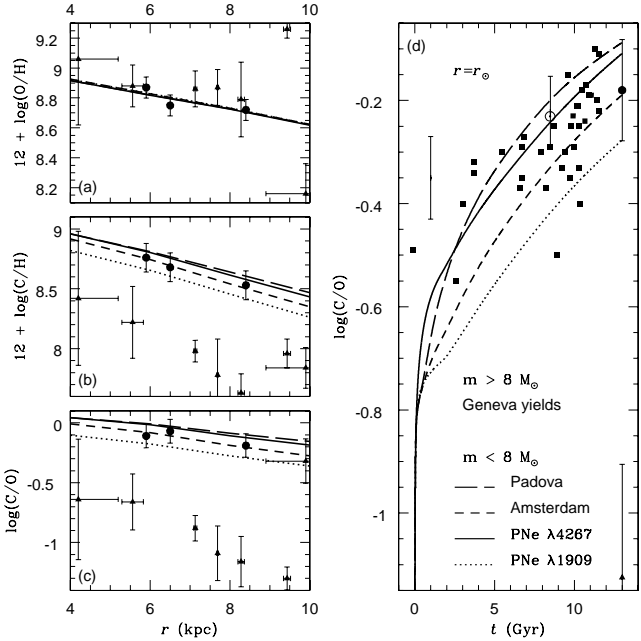} 
{\caption[]{
Predictions from models considering, for massive stars, yields by Maeder (1992) (Geneva yields)
and for low and intermediate mass stars,
yields by Marigo et al. (1996, 1998) and Portinari et al. (1998) (Padova),
van den Hoek \& Groenewegen (1997) (Amsterdam),
PN yields from forbidden lines (PNe $\lambda 1909$),
and from permitted line (PNe $\lambda 4267$).
(a)--(c) Present-day distribution of abundance ratios.
(d) The C/O evolution of the solar vicinity.
Observational data as in Figure 1.
}
\label{f3}}
\end{figure}

 All models are built to reproduce 
the observed gas fraction distribution of the Galaxy, $\sigma_{gas}/\sigma_{tot}$,
compiled by Matteucci \& Chiappini (1999) or Alib\'es, Labay \& Canal (2001)
and the observed O/H  Galactic gradient from 4 to 10 kpc.

The models are very similar to the infall model of Carigi (2000),
but 
in this work there are some differences in the assumptions about stellar yields:

a) Only two sets of metal-dependent stellar yields from massive stars
(MS, $8<m/{\rm M}_\odot<120$) are considered:
Geneva yields (Maeder 1992) and Padova yields (Portinari, Chiosi, \& Bressan 1998).
Carigi  concluded that
models with the Geneva yields or Padova yields can 
reproduce the increase of C/O with $Z$ in the solar vicinity,
but only models  with the Geneva yields can match the negative C/O gradient.
Models based on Woosley \& Weaver (1995) cannot fit the increase of C/O with $Z$
nor the negative C/O gradient. 

b) Four sets of stellar yields for low and intermediate mass stars
(LIMS, $0.8<m/{\rm M}_\odot<8$)  are used:
i) two  metal-dependent-theoretical ones:
the  Amsterdam yields (van den Hoek \& Groenewegen 1997)
and the Padova yields (Marigo et al. 1996, 1998,  and Portinari et al. 1998)
ii) two metal-independent-observational stellar yields:
PN yields from permitted  lines (PNe $\lambda4267$)  
and  PN yields from forbidden lines (PNe $\lambda 1909$).
For more details, see section 3.

c) For each set of yields
 linear interpolations for different stellar masses and metallicities
are made.

The evolution of the C/O ratio is determined by the number of stars that die and 
by the amount of C and O ejected (stellar yields)
 by each star to the interstellar medium. 
The number of stars is given by the SFR, the IMF and the lifetime of each star. 
Changes in the slope of the IMF modify the relative number of  massive stars
to LIMS and therefore modify the evolution of the C/O ratio.
Chemical evolution models of the Galactic disk have assumed different IMFs,
SFRs, and stellar properties 
(Tosi 1996, Carigi 1996,  Chiappini, Matteucci \& Gratton 1997,  Prantzos, Aubert \& Audouze 1996, 
Liang, Zhao \& Shi 2001, Alib\'es et al. 2002).
Based on those models it is not obvious to quantify the effect of each factor on the predicted abundances.

If the  SFR and the stellar yields are fixed, and
the IMF by  Kroupa, Tout \& Gilmore (1993) is replaced by the Salpeter IMF, 
C/O increases $\sim 0.10$ dex at late evolution and decreases $\sim 0.15$ dex 
at middle evolution (Carigi 2003).
Moreover, if a constant IMF is  replaced by 
a varying IMF with time or metallicity, some important observational constraints
in the solar vicinity cannot be reproduced (Carigi 1996, Chiappini, Matteucci, \& Padoan 2000). 

In all chemical evolution models of the Galactic disk
 the SFR is such that the distribution of the G-dwarf and Galactic gradients are reproduced.
A bursting SFR, (Chiappini et al. 1997,  Carigi, Col\I n, \& Peimbert 1999)
produces a decrease in C/O smaller than ~0.2 dex, immediately after each burst,
then C/O increases until it reaches a value similar to that obtained with a continuous SFR.

The variation of C/O with time or with O/H depends on
i) the C and O yields,
ii) the initial mass function, and
iii) the star formation rate.
Since the initial mass function (Kroupa et al. 1993),
the star formation rate
($ SFR \propto \ \sigma^{1.2}_{gas} \ \sigma^{0.2}_{gas+stars}$),
 and the massive-star yields are fixed, then
the C/O value can be used as a constraint to discriminate among the different sets of C and O yields for
low and intermediate mass stars.

\begin{table}
\caption{Carbon Ejected by  Stellar Populations during 13 Gyr}
\begin{center}
\begin{tabular}{lcccccc}
\multicolumn{2} {c} {Assumed Yields} & \hskip 0.1cm & \multicolumn{4} {c}{Contribution (\%)} \\
MS & LIMS           & & PNII/III & PNI  & SNII/Ib & SNIa \\
\hline
\hline
        & Padova              && 30.0 & 16.4 & 51.7 & 1.9 \\
        & Amsterdam           && 20.6 & 13.9 & 63.5 & 1.9 \\
Geneva  & PNe $\lambda4267$  &&   20.9 & 22.1 & 54.6 & 2.4 \\
        & PNe $\lambda1909$   && 15.5 &  6.4 & 76.0 & 2.1 \\
\hline
        & Padova               && 25.4 & 18.7 & 54.4 & 1.5\\
Padova  & PNe $\lambda4267$   && 18.3 & 22.9 & 56.9 & 2.0 \\
        & PNe $\lambda1909$   && 13.5 &  6.6 & 78.3 & 1.6 \\
\hline
\hline
\end{tabular}
\end{center}
\end{table}

\section{Results}
\label{sec:results}

\begin{figure}
\epsfxsize=9.0cm
\epsfbox{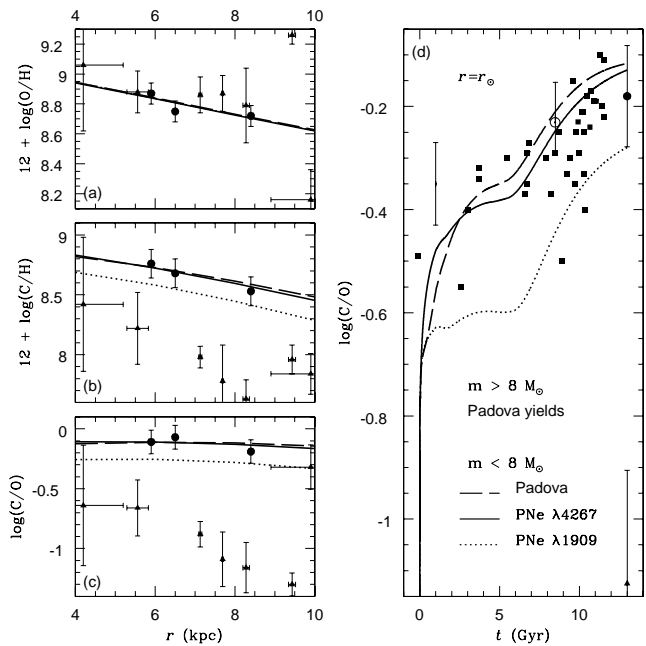} 
{\caption[]{
Same as Figure 3  but with
predictions from models considering yields
for massive stars by Portinari et al. (1998) (Padova yields)
and,
yields for low and intermediate mass stars
 by Marigo et al. (1996, 1998) and Portinari et al. (1998) (Padova),
 from forbidden lines (PNe $\lambda1909$)
and from permitted lines (PNe $\lambda4267$)
for low and intermediate mass stars.
}
\label{f4}}
\end{figure}

Predicted and observed C/H ratios in Planetary Nebulae are shown in Figure 2.
From this figure, it can be noted that:
i) the average C/H values from permitted lines are higher,  by as much as  a factor of 5,
than those from forbidden lines;
ii) there is a very good agreement between the average ${\rm(C/H)^{PL}_{PN}}$ value
  and those predicted
by stellar evolution models with $Z \ge Z_\odot$, for $3 < m/\msun < 8 $;  and
iii)  the average ${\rm (C/H)^{FL}_{PN}}$  yields  agree with those
computed 
from stellar evolution models with $Z \le 0.008$ for $3 < m/\msun < 8 $.
Otherwise, for $m < 3$ \msun \ the comparison between observational and
theoretical C/H values at different stellar metallicities and masses is
 neither clear nor simple.

The yields for objects with $m> 2.4$ \msun \ correspond to initial $Z$ values similar to the present ones.
This comparison is valid because the initial $Z$ values of PNe with $m> 2.4$ \msun
correspond to $Z$ values of population I objects, and the initial $Z$ values of PNe with
$m > 2$ \msun \ is expected to be higher than 0.01.

The effect that the observational yields produce on  the 
predicted abundances is difficult to estimate without the computation of detailed models. 
Therefore, we have computed chemical evolution models with observational yields to 
quantify that effect and
to try to discriminate between permitted line yields and forbidden line yields.
Predictions for the Galactic disk and the solar vicinity are summarized in Figures 3 to 5 
together with Tables 1, 2, and 3.

In Figure 3, I present predictions from those models that assume yields by Maeder (1992)
for stars with $m>8$ \msun. Carigi (2000) concludes that only with these yields 
the C/O Galactic  gradient and the C/O increase with time are  reproduced
and in this work the same conclusions are confirmed.
From this figure and  comparing  observed and predicted C/O abundances, it can be noted that:
i) The C/O values predicted with ${\rm C}^{\rm PL}_{\rm PN}$ yields are higher 
 than those
obtained with the Amsterdam yields and slightly lower 
(by as much as a factor of 0.05 dex)
than those computed with the Padova yields;
ii) Models based on theoretical yields and ${\rm C}^{\rm PL}_{\rm PN}$ yields 
reproduce both
the observed  C/O increase with time in the solar vicinity, the C/O gradient,
and the C/O values;
iii)
The C/O values predicted with ${\rm C}^{\rm FL}_{\rm PN}$ yields are lower than
those observed in the Sun and in most
dwarf stars located within 1 kpc of the Sun
(by factors of 0.05 and 0.1 dex, respectively);
iv)
 Models based on ${\rm C}^{\rm FL}_{\rm PN}$ yields
reproduce the C/O values
observed in   M17 and Orion, within 1 $\sigma$, while 
the M8 value deviates by 1.2 $\sigma$, which is a good agreement.

\begin{table}
\caption{Oxygen  Ejected by  Stellar Populations during 13 Gyr}
\begin{center}
\begin{tabular}{lcccccc}
\multicolumn{2} {c} {Assumed Yields} & \hskip 0.1cm & \multicolumn{4} {c}{Contribution (\%)} \\
MS & LIMS           & & PNII/III & PNI  & SNII/Ib & SNIa \\
\hline
\hline
        & Padova               && 6.4 & 4.8 & 86.3 & 2.5 \\
        & Amsterdam            && 6.2 & 4.7 & 86.7 & 2.5 \\
Geneva  & PNe $\lambda4267$   && 5.5 & 4.5 & 87.6 & 2.5 \\
        & PNe $\lambda1909$   && 5.4 & 4.5 & 87.7 & 2.5 \\
\hline
        & Padova               && 5.0 & 4.8 & 88.3 & 1.9 \\
Padova  & PNe $\lambda4267$   && 4.2 & 4.1 & 89.8 & 1.9 \\
        & PNe $\lambda1909$   && 4.1 & 4.1 & 89.9 & 1.9 \\
\hline
\hline
\end{tabular}
\end{center}
\end{table}

In Figure 4, I present  the predictions from models that consider yields by Portinari et al. (1998)
 for massive stars. Carigi (2000) concludes that models with these yields 
can reproduce the C/O increase with $Z$
in the solar vicinity, but not the C/O Galactic  gradient.
Both conclusions are confirmed in this work.
From this figure, the same results that are shown in Figure 3 can be noted
 among ${\rm C}^{\rm PL}_{\rm PN}$, ${\rm C}^{\rm FL}_{\rm PN}$,
and  theoretical yields,
but in this case, models based on the ${\rm C}^{\rm FL}_{\rm PN}$ yields 
deviate from the C/O values observed in the three \HII regions
by 1.3, 1.8 and 1 $\sigma$.

In Table 1 the predicted gradients at the present time are shown.
By comparing observed and predicted values, it can be noted that:
the C/H gradient is reproduced by all models, but the C/O gradient
is reproduced only by the models that assume Geneva yields, again confirming
the conclusion of Carigi (2000).

The C and O ejected (processed and not processed by stars)
 to the ISM during the whole evolution of the solar vicinity 
are presented in Tables 2 and 3.
Massive stars produce the greater part of O and eject between 52 and 64 \% of C.
LIMS do not synthesize O and eject  between 34 and 46 \% of C.

Liang et al. (2001),
concluded that in the late stage of evolution of the solar vicinity
LIMS and metal-rich Wolf Rayet stars eject an important amount of carbon,
but they were not able to distinguish which of these two groups of stars
 is the main source of carbon.
Based on our models, I conclude that  C enrichment in the late stage of galactic evolution
is mainly due to massive stars.

To analyze further the properties of the models I decided to use the
 C/O versus O/H diagram.
In Figure 5, I present  the solar vicinity models for the seven sets of yields
used in this paper. 

 I have divided the evolution of the solar vicinity in three stages: 
early ($t< 0.5$ Gyr, log (O/H)$ <  -6.2$ dex),
middle ( 0.5  Gyr$ < t <$ 5.5 Gyr, $-6.2$ dex $<$ log (O/H)$ < -3.7$ dex),
and late ($t >$ 5.5  Gyr, log (O/H)$ >  -3.7$ dex).  
In the early stage  C is produced only by massive stars and their contribution
to the  C/O ratio is low;
in the middle stage the C production is due to  LIMS and massive stars, both kinds
of stars have
C yields that depend on their initial O/H values;
in the late stage the C/O increase with O/H is determined by metal-rich massive stars.  

For comparison, I have included the data presented by Nissen (2002, Figure 4).
In that figure he presents [C/O] and [O/H] values:
i) in halo main sequence stars from high excitation lines in  near-IR VLT/UVES spectra,
ii) in halo stars from high excitation lines by Tomkin et al. (1992),
iii) in disk stars from forbidden lines with 3D model atmosphere corrections.
I have adopted the same solar abundances that Nissen and Tomkin et al. assumed,
(Allende et al. 2001, 2002) and Grevesse et al. (1991), respectively,
 to obtain the abundance ratios shown in Figure 5.

\begin{figure}
\epsfxsize=9.0cm
\epsfbox{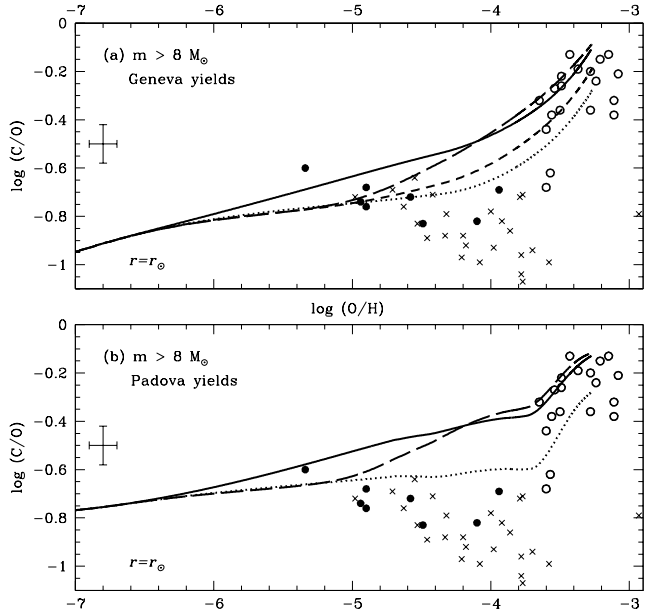} 
{\caption[]{
Log(C/O) -- log(O/H) relation for $r=r_\odot$.
Predictions of all models presented in Figures 3 and 4 that assume
(a)  Geneva yields or (b) Padova yields for massive stars.
The symbols used for the relations are the same as those
presented in Figures 3 and 4, respectively.
Observational data from several sources revised by Nissen (2002). Open circles: disk stars. Filled circles:
halo stars, VLT/UVES data, Nissen (2002). Crosses: halo stars, Tomkin et al. (1992).
Error bars at the left represent the typical error. 
}
\label{f5}}
\end{figure}

In Figure 6, I present Galactic models for two galactocentric distances, $r=4$ and 10 kpc,
for only two sets of yields, C$_{\rm PN}^{\rm PL}$ and C$_{\rm PN}^{\rm FL}$ yields,
and compare the model results with observations of \HII regions in the Galaxy,
M101, NGC 604, and NGC 2363.
All the measurements shown were derived
from abundances determined from PL.
Notice that for each model the only C/O value that corresponds to
the present time is the highest O/H one, the other O/H values correspond to
earlier times.
For $r> 10$ kpc the present-time C/O and O/H values are lower than
the final values shown for $r=10$ kpc.

From  Figure 6 it follows that
 the Geneva yields (panel a) together with  C$_{\rm PN}^{\rm PL}$ yields 
provide a good fit to NGC 5471, a giant \HII region in M101
(a ScdI galaxy), to NGC 604, the brightest \HII region in M33 (Sc,cdII-III)
and to Orion and M8 in the Galaxy (SbcI-II);
while for the Padova yields (panel b) together with
  C$_{\rm PN}^{\rm PL}$ yields the fit for these
objects is only fair.
On the other hand, for these objects the Geneva and Padova yields together with 
C$_{\rm PN}^{\rm FL}$ yields produce a poor fit.

For NGC 5461, the other \HII region in M101,
the Esteban et al. (2002) value 
(C/O$=-$0.4 dex; O/H$=-$3.1 dex) is not fitted by PL or FL models.
Garnett et al. (1999) obtain C/O $=-$0.20 dex 
and O/H$=-$3.49 dex for Av=4.1. 
The large differences between these two sets of abundances probably indicate
that observations of higher quality are needed for this object.

It is also apparent from Figure 6 that NGC 2363, a \HII region in NGC 2366 and 
the most metal-poor \HII regions of the sample,
can be fitted with Geneva or Padova yields 
with the observational C$_{\rm PN}^{\rm FL}$ yields.
This result probably implies that
the accuracy of the
 C/O determinations for this object is  not yet good enough to be able to distinguish among
the different models
or
 the model for the Galaxy does not apply to
NGC 2366 (Irr).
This \HII region might be best fitted with models
tailored to irregular galaxies (eg. Carigi et al. 1995, 1999, van Zee et al. 1998)

For the models based on Geneva yields 
the  C/O gradient steepens with time, because the C yields increase with $Z$
and the O yields decrease with $Z$ for massive stars.
But for the models based on the Padova yields 
the gradient flattens in the last stage evolution
because the C/O values of the material ejected by  supermetallic massive stars
and  metal-poor stars are similar.
From Figure 6 it can be seen that at present (the highest O/H values for
each model) the Geneva gradients are steeper than the Padova ones.

The Milky Way is the most evolved galaxy of our sample,
therefore its gas content is lower and its \HII regions
are more metal-rich.
M101 is less evolved, its gas content is higher and its \HII regions are less metal-rich.
To understand better the evolution of M101,
it is necessary to compute models tailored to fit  this galaxy.

\section{Discussion}
\label{sec:discussion}

The C evolution of the solar vicinity and the Galactic disk evolution are well
understood when theoretical yields (Amsterdam or Padova),
which are dependent on $Z$ and $m$, are assumed.

It is complicated to predict the effect of the observational yields,
which are independent of $Z$ and $m$,
on the C evolution, since
i) the initial mass function predicts more low-mass stars than intermediate mass stars
and
ii) the low mass stars 
contribute to the gas enrichment more slowly than
the intermediate mass stars.

Therefore,
we have computed chemical evolution models
based on a successful model for the Galactic disk
 with observational yields from C/H abundances in PNe
to try to discriminate between
the yields derived from permitted lines and those derived from forbidden lines.

Models that assume  observational yields
from 
permitted lines can reproduce the C/O ratios observed in \HII regions
and stars of the solar vicinity,
but models that consider yields from
forbidden lines cannot.

The difference of 0.7 dex between the average (C/H)$^{\rm PL}_{\rm PN}$ and
(C/H)$^{\rm FL}_{\rm PN}$ ratios is echoed by the predicted  C abundance, 
showing a difference of 0.15 dex 
between C/H ratios predicted by models that consider C$^{\rm PL}_{\rm PN}$  and
C$^{\rm FL}_{\rm PN}$ yields.

From Figure 5,
it can be seen that the halo stars observed 
with VLT/UVES by Nissen (2002) exhibit a nearly
constant C/O value of $-0.75$ dex, 
while the mean  of the measurements of Tomkin et al. (1992) 
is log C/O $\simeq -0.85$.
Moreover, models based on massive star yields by Maeder (1992) predict C/O values lower than those
with yields by Portinari et al. (1998). Therefore, 
models with Maeder yields and theoretical yields for LIMS reproduce the C/O values observed in halo stars,
within 1 $\sigma$ for O/H $<-4.6$ and within 3 $\sigma$ at O/H $\sim -4$, a fair agreement.
But models with  Portinari et al. yields for massive stars and Padova yields for LIMS
deviate from the C/O values observed in the halo stars, 
by 1 $\sigma$ for O/H $<-5.0$ and  4 $\sigma$ at O/H $\sim -4$, a poor agreement.

Carigi (2000) suggests the yields of massive stars increase with $Z$ and  one of the issues of this work
is to find if a similar dependence exists in LIMS yields.
The C/O values of halo stars test   the models at low $Z$, while the PNe test the models at high $Z$.
From Figure 5 it follows that the yields with constant $Z$ can adjust objects with O/H values
$>-3.6$ dex, but not the halo stars with O/H values $<-3.6$ dex. 
To adjust the halo stars it is necessary to assume that the yields increase with $Z$.
The models with theoretical yields provide a reasonable fit to the observations by Nissen (2002), but do not
fit the results by Tomkin et al. (1992)
Since the ${\rm C}^{\rm FL}_{\rm PN}$ models are closer to the halo data than the 
${\rm C}^{\rm PL}_{\rm PN}$ models it follows that C yields at lower metallicities 
should be lower than those at higher metallicities, 
which also supports the result that the yields should increase with $Z$.
It is not possible to observationally estimate the C yields for LIMS from
halo PNe, because they originate from progenitor stars in the 0.8 to 1.0 \msun range only.

It possible to predict the future evolution of the  solar vicinity, and it is very similar
to the final evolution for $r=4$ kpc shown in Figure 6.
According to the data set by Nissen (2002), C/O should level off to a plateau at O/H $>-3$ dex, and
that plateau is predicted by both sets of models.
In fact,  with the Padova yields, the models reach a plateau  more quickly, but these models predict
flat C/O gradients.
Moreover, data by Nissen (2002) show a steep rise of C/O around O/H $\sim -3.6$ dex,
which is better reproduced by models with the Padova yields than those with the Geneva yields.

To discriminate among  models
in the early and early-middle evolutionary stage,
it is necessary to use as constraints C/H and O/H values of very metal-poor objects.
Unfortunately, C abundances in damped Ly$_\alpha$ systems (DLAs) at high
redshifts (possible protospiral galaxies) have so far been proved
difficult to determine (e.g. Prochaska \& Wolfe 2002).
Recently, L\'opez et al. (2002) have estimated a lower limit for a
possible dust-free DLA at $z=2.3$, ([C/H]\,$> -1.43$, [O/H]\,$=-0.81$), 
but this C/O value is not accurate enough to constrain the models.
 
It would be important to determine C and O abundances of spiral galaxies 
at different distances to test the models.
According to the models the   
C/O gradient would have a bimodal behaviour with $z$:
the gradient gets steeper with $z$ at middle and high redshifts,
but then, it flattens out at very high $z$.

\begin{figure}
\epsfxsize=9.0cm
\epsfbox{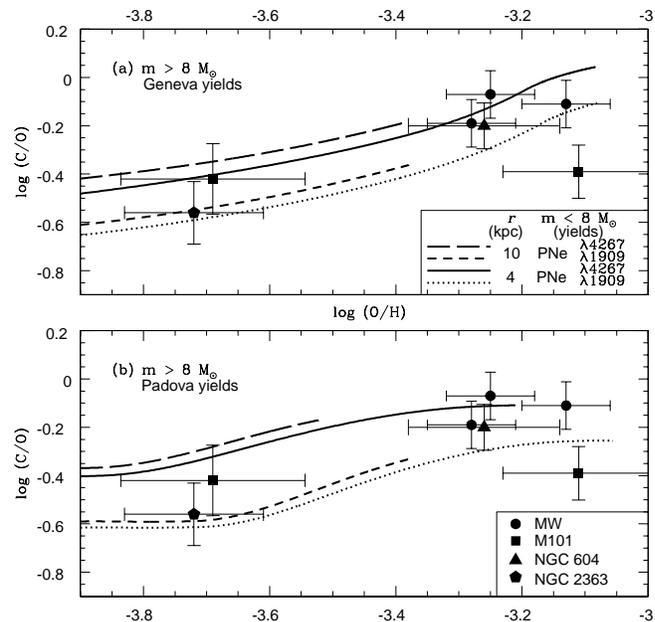} 
{\caption[]{
Late log(C/O) -- log(O/H) relation for $r=4$ and 10 kpc.
Predictions for models that assume PN observational yields for LIMS and (a)
Geneva yields or (b) Padova yields for massive stars.
Observational data for Galactic \HII regions from
 Esteban et al. (1998, 1999, 2002)
and for  extragalactic \HII regions  from Esteban et al. (2002).
A 40 \%  uncertainty is assumed for values of Esteban et al. (2002)
in cases where they do not quote an error.
Corrections for dust depletion according to  Esteban et al. (1998). 
}
\label{f6}}
\end{figure}

\section{Conclusions}
\label{sec:conclusion}

From chemical evolution models of the Galaxy,
I conclude that:
 
a) Models with the permitted line yields (C$_{\rm PN}^{\rm PL}$)  
 match all the observational constraints,
in particular  they reproduce
the C/O absolute values observed in dwarf stars of the solar vicinity
and in \HII regions of the Galactic disk.
 
b) Models with the forbidden line yields (C$_{\rm PN}^{\rm FL}$)  
fail to  reproduce the
C/O ratios in dwarf stars of different ages  in the solar vicinity,
the Sun, and the inner \HII region M17.
 
c) Models with C$_{\rm PN}^{\rm PL}$ yields 
agree with models based on theoretical yields,
in particular showing better agreement
with models based on the Padova yields  than
models based on the Amsterdam yields.
 
d) The C/O values predicted with ${\rm C_{PN}^{\rm PL}}$ yields
 are about 0.08 dex higher  than those
obtained with the Amsterdam yields, and about 0.03 dex lower
than those computed with the Padova yields.

e) The C/O values predicted with ${\rm C_{PN}^{\rm FL}}$ yields
 are about 0.10 dex lower than those
obtained with the Amsterdam yields, and about 0.20 dex lower
than those obtained with the Padova yields.

f) The C$_{\rm PN}^{\rm PL}$ yields 
should increase with $Z$ to obtain a better 
 agreement between models and observations in the
 C/O versus $t$ and  C/O versus O/H diagrams.

g) The C/O increase with $Z$ is governed by the metallicity dependent yields
of both massive stars and LIMS. 
Massive stars determine the behaviour of C/O with $Z$ in the early and late evolution,
while LIMS do so in the middle evolution.

h) The C/O gradient steepens with time, but when the gas acquires supersolar abundances,
the gradient flattens with time.

\section*{\bf ACKNOWLEDGMENTS}
I dedicate this paper to 
Manuel Peimbert and Silvia Torres-Peimbert.
I am grateful to Poul E. Nissen for providing me the VLT/UVES data
prior to publication and for fruitful discussions.
I wish to thank the referee for his excellent suggestions
which helped to improved and clarify this final version.
I also acknowledge a thorough reading of the manuscript by William Lee.
I thank the Instituto de Astrof\I sica de Andaluc\I a,
where part of this paper was written,
for its hospitality.

\end{document}